\documentstyle[12pt,preprint,prd,aps,amsfonts]{revtex}
\tighten
\begin{document}
\title{Approximate spectral functions in thermal field theory\thanks{
 Work supported by GSI}}
\author{P.A.Henning
       \thanks{e-mail: P.Henning@gsi.de},
        E.Poliatchenko and T.Schilling}
\address{Theoretical Physics, Gesellschaft f\"ur Schwerionenforschung (GSI) \\
        Planckstra\ss e 1, D-64291 Darmstadt, Germany }
\author{J.Bros}
\address{CEA, Service Physique Th\'eorique, CE-Saclay \\
        F-91191 Gif-sur-Yvette Cedex, France}
\date{ May 1996 (revised)}
\maketitle
\begin{abstract}
Causality requires, that the (anti-) commutator of two
interacting field operators vanishes for space-like
coordinate differences. This implies, that the
Fourier transform of the spectral function of this
quantum field should vanish in the space-like domain.
We find, that this requirement imposes some constraints on the use
of resummed propagators in high temperature gauge theory. 
\end{abstract}
\pacs{05.30.Ch, 11.10.Wx, 12.38.Mh, 25.75.+r, 52.25.Tx}
\section{Introduction}
In the past decade the interest in quantum field theory at 
nonzero temperature has grown considerably \cite{CJP93,banff93,china95}. 
In part this is due to experimental and theoretical efforts to understand
various hot quantum systems, like e.g. ultrarelativistic heavy-ion collisions
or the early universe.

A particularly interesting problem when investigating such systems is the  
question of their single particle spectrum. Although for an 
interacting quantum field this spectrum in general has a very rich structure, 
we may also count calculations of particle masses and spectral width 
parameters (i.e., damping rates) in this category.
Consequently we find, that a large number of research papers is dealing with
the question of the single particle spectrum on an approximate level.

With the present paper we address the question, whether the approximations
used in many of these papers are consistent with basic requirements 
of quantum field theory. To this end we investigate some common
approximations made to the quantity which summarizes the spectral properties
of a quantum field, i.e., its {\em spectral function\/} 
${\cal A}(E,\bbox{p})$. Up to a factor this function is the 
imaginary part of the full retarded two-point function, propagating 
an excitation with energy $E$ and momentum 
$\bbox{p}$. We find indeed, that some approximations to the spectral
functions require great care in their usage. 

The paper is organized as follows: First we present a 
brief introduction dealing with free boson and fermion quantum fields.
We then investigate the properties of simple approximate
spectral functions for interacting boson and fermion fields, and
finally we turn to the physical problem of gauge theory at finite temperature.
\section{Simple spectral functions and commutators}
The spectral function has two features which are intimately related to 
fundamental requirements of quantum field theory.
To begin with, the quantization rules for boson and fermion fields 
(which we will use in the free as well as in the interacting case),
\begin{eqnarray} \nonumber
\left[\phi(t,\bbox{x}),\partial_t \phi(t,\bbox{y})\right] &=& {\mathrm i}
  \delta^3(\bbox{x}-\bbox{y})\;,\\
\label{ccr}
\left\{\psi(t,\bbox{x}),\psi^\dagger(t,\bbox{y})\right\} &=& 
  \delta^3(\bbox{x}-\bbox{y})
\;,\end{eqnarray}
require that the spectral function is {\em normalized\/}. Although 
the normalization may be difficult to achieve numerically,
we do not consider this a serious principal problem. 

The second important feature of the spectral function is that
its four dimensional Fourier transform into coordinate space must vanish 
for space-like arguments. This is equivalent to the Wightman axiom of 
{\em locality\/}, i.e., field operators must
(anti-)commute for space-like separations in Minkowski space \cite{GJ81}.

In an interacting many-body system, we may very well expect 
non-locality in a causal sense: Wiggling the system at one side will
certainly influence the other side after some time. The 
locality axiom ensures that this influence does not occur over
space-like separations, i.e., faster than a physical signal
can propagate. Thus, to distinguish between the {\em causal\/} non-locality 
and the violation of the locality axiom, we will henceforth  
denote the latter a violation of {\em causality\/}.

In the following we will furthermore distinguish fermionic and bosonic
quantities by a lower index.  For two field operators the locality
axiom then amounts to the requirement, that the (anti-) commutator
function of two field operators and also its expectation value
fulfills
\begin{equation}
\left.{\array{rrr}
C_B(x,y) &=& \vphantom{\int\limits_0^0}\left\langle\left[\phi(x),
  \phi(y)\right]\right\rangle\\
C_F(x,y) &=&\vphantom{\int\limits_0^0}\left\langle\left\{\psi(x),
  \psi^\dagger(y)\right\}\right\rangle\endarray}\right\}
= 0\;\; \mbox{if}\;\;(x-y) \;\mbox{space-like}
\;.\end{equation} 
In terms of the spectral function, these expectation values are
\begin{equation}\label{cdef}
C_{B,F}(x-y) = \int\!\!\frac{dE\,d^3\bbox{p}}{(2\pi)^3}
  \;{\mathrm e}^{\displaystyle -{\mathrm i}(E(x_0-y_0)-
  \bbox{p}(\bbox{x}-\bbox{y}))}\,{\cal A}_{B,F}(E,\bbox{p}) 
\;.\end{equation}
We first consider the case of free quantum fields, for completeness
we quote the free spectral functions found in any textbook on field theory:
\begin{eqnarray}\nonumber
{\cal A}^0_B(E,\bbox{k}) &=&
        \mbox{sign}(E)\,\delta(E^2-\bbox{k}^2 - m_B^2)\\
\label{ff}
{\cal A}^0_F(E,\bbox{p}) &=&
        \left( E\gamma^0 + \bbox{p}\bbox{\gamma} + m_F\right)\,
        \mbox{sign}(E)\,\delta(E^2-\bbox{p}^2 - m_F^2)
\;.\end{eqnarray}
For these spectral functions one obtains as the (anti-) commutator 
expectation value 
$C^0_B(x,y) = C_0(x-y)$ and $C^0_F(x,y) = \left(
{\mathrm i}\gamma^\mu\partial_\mu + m_F\right) C_0(x-y)$, where
\begin{equation}\label{cfree}
C_0(x) = \frac{ - {\mathrm i}}{2\pi}\;
  \left( \delta(x_0^2-\bar{x}^2) - \Theta(x_0^2-\bar{x}^2)
  \frac{m_{B,F}}{2 \sqrt{ x_0^2-\bar{x}^2 }}\,
  J_1(m_{B,F} \sqrt{ x_0^2-\bar{x}^2}) \right)
\;.\end{equation}
Clearly this is zero for space-like arguments, i.e., for
$\bar{x} = |\bbox{x}| > |x_0|$. This free commutator function has
support only in the unshaded area of Fig. 1, and it is singular at its
boundaries (but zero outside).

We now turn to nontrivial spectral functions, which are more
appropriate for a thermal system. The physical reason is, that at
finite temperature particles are subject to collisions, hence their
state of motion will change after a certain time. In a hot quantum
system therefore the {\em off-shell\/} propagation of particles plays
an important role. This off-shellness is contained in a continuous
spectral function, which must not have an isolated $\delta$-function
like pole.  In principle this means that at nonzero temperature {\em
every\/} quantum system must be described on the same footing as a gas
of resonances. However, this does not imply that thermal particles may
decay -- they are merely scattered thermally by the other components
of the system.

Apart from this physically motivated use of continuous spectral
functions at finite temperature, one may also adopt a mathematically
rigorous stance. We do not elaborate on this, but rather quote the
Narnhofer-Thirring theorem \cite{NRT83}. It states,
that interacting systems at finite temperature {\em cannot\/} be
described by particles with a sharp dispersion law, only
non-interacting ``hot'' systems may have a $\delta$-like spectral
function. Ignoring this mathematical fact one finds as an echo serious
infrared divergences in high temperature perturbative quantum
chromo dynamics (QCD). Consequently, these unphysical singularities are
naturally removed within an approach of finite temperature field
theory with continuous mass spectrum \cite{L88,h94rep}.

Thus, for a mathematical as well as a physical reason, finite
temperature spectral functions are more complicated than those given
in eq. (\ref{ff}). The question then arises, how much more complicated
they have to be in order to be consistent with the requirements we
have discussed above: Fully self consistent calculations of the
corresponding spectral functions are very rare due to the numerical
difficulties involved \cite{h94rep,h92fock,KM93}. More often one uses
an {\em ansatz\/} for such a function which involves only a small
number of parameters which are then determined in a more or less
``self''-consistent scheme.

As an example we consider two seemingly simplistic generalizations of the
spectral functions in eq. (\ref{ff}), which involve only one additional
parameter:
\begin{eqnarray}\nonumber
{\cal A}^1_B(E,\bbox{ k}) 
&=&\frac{1}{\pi} \vphantom{\int\limits_0^0}
\,\frac{2 E \gamma_B}{\left(E^2-\bbox{k}^2 - m_B^2 - \gamma_B^2\right)^2+
4 E^2 \gamma_B^2}\\
\nonumber
{\cal A}^1_F(E,\bbox{p}) &=& \frac{\gamma_F}{\pi} \vphantom{\int\limits_0^0} 
\frac{\gamma^0\left(E^2 + \omega(\bbox{p})^2+ \gamma^2_F\right) +
      2 E \bbox{\gamma}\bbox{p} + 2 E m_F}{
  \left(E^2 - \omega(\bbox{p})^2 - \gamma^2_F\right)^2 + 4 E^2 \gamma_F^2}
\\ \nonumber
 &=&\frac{1}{4\pi{\mathrm i}\omega(\bbox{p})} \left(
\frac{\omega(\bbox{p})
  \gamma^0 + \bbox{p}\bbox{\gamma} + m}{E - \omega(\bbox{p}) - 
   {\mathrm i} \gamma_F}
-\frac{-\omega(\bbox{p})\gamma^0 + \bbox{p}\bbox{\gamma} + m}{
E+\omega(\bbox{p})
 - {\mathrm i} \gamma_F}\right.
\\ \label{sfan}
 &&\hphantom{\frac{1}{4\pi{\mathrm i}\omega(\bbox{p})}} \left.
-\frac{\omega(\bbox{p})\gamma^0 + \bbox{p}\bbox{\gamma} + m}{
E - \omega(\bbox{p})
 + {\mathrm i} \gamma_F}
+\frac{-\omega(\bbox{p})\gamma^0 + \bbox{p}\bbox{\gamma} + m}{
E+\omega(\bbox{p})
 + {\mathrm i} \gamma_F}
\right)\label{af}
\end{eqnarray}
where $\omega(\bbox{p})^2=  \bbox{p}^2 + m_F^2$. These relativistic
Breit-Wigner functions are somewhat oversimplified as compared with
the real world: They attribute the same spectral width to very fast
and very slow particles. Indeed, even when {\em approximating\/}
a more sophisticated calculation of a spectral function by 
simple poles in the complex energy plane, one obtains a 
strongly momentum dependent spectral width parameter $\gamma$
(see \cite[pp. 350]{h94rep} for an example).

However, for some physical effects the influence of fast particles is
reduced by Bose-Einstein or Fermi-Dirac distribution functions, such that 
one may use these simple spectral functions. Their constant spectral
width parameters $\gamma$ then may be considered as a parametrisation
of the dominant low-energy phenomena. A good example for such
a physical effect is the radiation of soft photons from a hot
plasma, i.e., the ``glow'' of the plasma, where the
ansatz of a constant spectral width parameter gives
results comparable to the classical Landau-Pomeranchuk-Migdal effect
\cite{hq95gam}.

It is a matter of a few lines to show, that the (anti-) commutator 
functions for the quantum fields defined by these spectral functions are
\begin{eqnarray}\label{cone}
\nonumber
C_B^1(x,y) &=&\vphantom{\int\limits_0^0} {\mathrm e}^{\displaystyle -\gamma_B
 \left|x_0-y_0\right|}\,C_0(x-y)\\
C_F^1(x,y) &=& \vphantom{\int\limits_0^0} 
  {\mathrm e}^{\displaystyle -\gamma_F
 \left|x_0-y_0\right|}\,
 \left({\mathrm i}\gamma^\mu\partial_\mu + m_F\right)\,
 C_0(x-y)
\;.\end{eqnarray}
These simple generalizations of the free spectral function therefore
have the important property to preserve causality: Their Fourier
transform vanishes for space-like arguments. Consequently, if these
spectral functions are used to construct a generalized free field
theory \cite{L88,h94rep}, it will be local.

Let us note at this point, that 
the most general form of a spectral 
function which conforms with this requirement 
has been given in ref. \cite{BB92}. We are not, however, interested
in the most general spectral function, but in those which are only
{\em slightly\/} more complicated than the free case.
\section{Hot gauge theory}
To this end, we turn to study hot gauge theory, as discussed in the
current literature \cite{W82,P89,W89,BI95}. Naturally we cannot possibly check
all the existing calculations of spectral functions, and therefore
restrict ourselves to the most basic picture obtained in high-temperature
QED. Up to a single difference this exactly comprises the
fermion gauge boson spectral functions obtained in the hard thermal loop
resummation scheme of QCD \cite{P89}. 

Let us first study the fermion of this model, which has a propagator
\cite{W89} 
\begin{eqnarray}\nonumber
&&S(E,\bbox{p}) =\\ 
\label{sfer}
&&\left[ E\gamma^0\,\left(1-\frac{M^2_D}{2 E \bar{p}}\,
 \log\left(\frac{E+\bar{p}}{E-\bar{p}}\right)\right)
-
\bbox{p}\bbox{\gamma}\left(1+\frac{M^2_D}{\bar{p}^2}\,
 \left(1-\frac{E}{2\bar{p}}\,
 \log\left(\frac{E+\bar{p}}{E-\bar{p}} \right)
\right)\right) \right]^{-1}
\;.\end{eqnarray}
Here, $\bar{p} = |\bbox{p}|$, and
$M_D$ is the Debye screening ``mass'', proportional to the temperature.
The spectral function of this propagator has a rather complicated
structure, described in detail in ref. \cite{W89}: Four discrete poles
at energies $\pm E_p$ and $\pm E_h$ with $E_p,E_h > \bar{p}$ on the
real axis; and a continuum for
$-\bar{p}<E<\bar{p}$. Each of these pieces contributes to the Fourier 
transform as may be seen from the top panel of Fig. 2. 

The four dimensional Fourier transform is a linear functional of
the imaginary part of the propagator. Thus, each contribution 
to the spectral function may be transformed separately, and their sum then
constitutes the total Fourier transform.

It is a priori clear that this total Fourier transform must be 
in agreement with the locality axiom. This follows from the fact, that
$S$ is holomorphic in the forward tube, i.e., for timelike
imaginary part of the four vector $(E,\bbox{p})$. However,
only the sum of all contributions vanishes in the spacelike region,
and therefore locality and causality are only guaranteed if spacelike 
{\em and\/} timelike four-momenta are taken into account in the
propagator (\ref{sfer}).
A restriction in the fashion $|E| < \bar{p}$ leads to a violation
of causality.
 
It was already mentioned that there exists a difference between the
hard thermal loop resummation (HTL) scheme and high temperature QED (or QCD).
In the HTL method, the ``dressed'' propagators are used only for soft 
momenta which are smaller than $\sqrt{8}\,M_D$, for high momenta one is
required to use free propagators. The region of intermediate momenta 
is usually ignored in this method.

The locality axiom provides a convenient method to check for the
validity of this approximation. To this end, we ``patch'' the 
free and resummed propagator together at the separation scale
$\bar{p}=\sqrt{8}\,M_D$.

In the bottom panel of Fig.2 we show the anti-commutator
function of two fermion fields using this prescription. Clearly, even
the sum of all contributions does not vanish outside the physical
region. We therefore conclude at this point, that no local quantum
field theory can be constructed which conforms to the ``patching''
rule for the propagators. Consequently, one may not ignore the
intermediate momentum region in hot gauge theory.

In the next step, we consider the gauge boson propagators, which for
transverse and longitudinal degrees of freedom are
\begin{eqnarray}\nonumber
\Delta_t(E,\bbox{p})& =&\left[ 
  E^2 - \bar{p}^2 - q^2_D\,\left( \frac{E^2}{2 \bar{p}^2}
  + \frac{E (\bar{p}^2 - E^2)}{4 \bar{p}^3}\, 
  \log\left(\frac{E+\bar{p}}{E-\bar{p}}\right)\right)\right]^{-1}\\
\label{hqed}
\Delta_l(E,\bbox{p})& =& \left(\frac{\bar{p}^2}{E^2 - \bar{p}^2}\right)\;\;
  \left[ \bar{p}^2 + q^2_D\,\left(1 - \frac{E}{2 \bar{p}}\,
  \log\left(\frac{E+\bar{p}}{E-\bar{p}}\right)\right)\right]^{-1}\;.
\end{eqnarray}
$q_D$ is the bosonic Debye screening ``mass'', which is proportional 
to the plasma frequency; it sets the only scale inherent to these
propagators.

Both of them have a continuous imaginary part (= spectral function)
in the regime $|E| < \bar{p}$, as well as a $\delta$-function
pole at some energy $> \bar{p}$. The analytical structure of these
propagators is quite complicated, but similar to the fermionic case it
may be shown that they are holomorphic functions
in the forward tube, and therefore their total Fourier transform is
zero outside the physical region. 

However, a
restriction to spacelike momenta, i.e., to
$|E|<\bar{p}$, leads to a violation of causality. A similar statement
holds if these propagators are used only for timelike momenta, and
consequently one should not consider plasmon propagation separately from
``collective'' effects.

The full gauge boson propagator is a linear combination of transverse
and longitudinal piece with certain projection factors and a gauge parameter
$\alpha$. In particular, the canonical 33-component is
\begin{equation}\label{d33}
\Delta_{33}(E,\bbox{p})
 = -\left(\vphantom{\int\limits_0^0} 
1 - \frac{p_3 p_3}{\bbox{p}^2}\right)\,\Delta_t(E,\bbox{p})
   -\frac{E^2 \,p_3 p_3}{\bbox{p}^2(E^2-\bbox{p}^2)}\,\Delta_l(E,\bbox{p})
+ \alpha\frac{p_3 p_3}{(E^2-\bbox{p}^2)^2}
\;.\end{equation}
We therefore have to study the Fourier transform of these products,
and thereby concentrate on the transverse piece since its
Fourier transform is numerically easier to obtain.

In the two panels of Fig. 2 we show the Fourier transform of the 
continuous part (spacelike momenta) and the plasmon part (timelike
momenta) of the transverse piece of $\mbox{Im}(\Delta_{33})$,
i.e., of $(p_3^2/\bbox{p}^2-1)\Delta_t$. The plot was made
for several values of $\bar{x}=|\bbox{x}|$ as function of $t$ (See Fig. 1
for the location of the displayed curves in the $x$-$t$ plane).
To each curve in the figure, we have added a thin vertical line
separating the regions inside and outside the forward lightcone. 

In the (shaded) region outside the forward cone, 
the two different contributions have the same sign and do 
{\em not\/} cancel each other in the total Fourier transform.
It now remains the question, whether this is cured by taking into account
the longitudinal piece of the propagator $\Delta_{33}$.

Obviously, since $\Delta_t$ and $\Delta_l$ are local by themselves, the 
violation of locality we saw above is due to the projection factors. 
Specifically it is due to the factor $p_3p_3/\bbox{p}^2$ which introduces 
a branching-point singularity at $\bbox{p}=0$ in the forward tube.
However, as is easily noted,
\begin{equation}
\lim_{\bbox{p}\rightarrow 0} \Delta_t(E,\bbox{p}) =
\frac{1}{E^2-\frac{1}{3}q^2_D} =
\lim_{\bbox{p}\rightarrow 0} \Delta_l(E,\bbox{p})
\;,\end{equation}
which implies that the branching point singularities cancel in the
sum of transverse and longitudinal piece of eq. (\ref{d33}).
We therefore term this violation of the locality axiom a purely {\em
kinematical\/} one, which is cured by using transverse {\em and\/}
longitudinal propagator on the same footing.

Consequently, the canonical gauge boson 
propagators $\Delta_{\mu\nu}(E,\bbox{p})$
of hot gauge theory are holomorphic functions for timelike imaginary part 
of $(E,\bbox{p})$, i.e., in their forward tube. Their Fourier transform
vanishes for $|t|< \bar{x}$, and therefore they obey the locality axiom.

However, as pointed out before, this necessitates the use of the
resummed propagator for all momenta. If for ''hard'' momenta
$\bar{p}> \sqrt{3} q_D$ they are replaced by the free boson
propagator, or forcibly set to zero,
causality is violated. From a more mathematical viewpoint, the
``patching'' of propagators breaks the principle of analytical
continuation. The violation of locality arises, because the ``patched''
propagators are no longer globally holomorphic in the forward tube.

Similarly, causality is violated if the propagators are restricted to
spacelike or timelike momenta alone, as may be seen from Fig. 2 and 3.
Another problem of remains even if the resummed propagators
are used globally, i.e., for soft as well as for hard momenta: They do
not conform with the relativistic Kubo-Martin-Schwinger boundary
condition, which requires an exponential falloff in the high-momentum
limit \cite{BB94}.
\section{Conclusion}
One may draw three conclusions from the present work. Firstly we
find, that seemingly simplistic {\em ansatz\/} spectral functions as
given in eqs. (\ref{af}) obey the axiom of locality, i.e.,
they allow only causal non-locality. This makes them a reasonable starting
point for any nonperturbative treatment of matter at high temperature.

The second conclusion is associated with the perturbative treatment 
of particles in hot gauge theory. Let us first discuss, whether
a possible violation of locality in this case
is of any relevance for measurable
quantities: Following ref. \cite{P89} one may argue, that the gauge
field itself has no physical meaning. However, the commutator of two
magnetic field components in our example, where the
commutator expectation value of different space-like components
is zero, reads  
\begin{equation}
\left\langle\left[B_i(x),B_j(y)\right]\right\rangle = 
 \varepsilon^{ijk}\,\frac{\partial^2}{\partial x_i \partial x_j}\,
\left\langle\left[A_k(x),A_k(y)\right]\right\rangle
\;.\end{equation}  
This implies, that in the present example a non-vanishing commutator
function of the gauge field outside the light cone generally leads to
a non-vanishing commutator function for observable quantities.
However, this statement has to be restricted: The purely
kinematical violation of causality we observed when not combining
transverse and longitudinal degrees of freedom will be canceled by 
moving to magnetic fields.

If we exclude this case, the violation we are discussing here would have the
physical effect that the magnetic field could not be ``measured''
independently at two points with a space-like separation.

Another example is the electric field at space-time coordinate
$(t,\bbox{x})$ produced by a transverse $\delta$-function perturbation
$\propto \delta^4(y)$ at space-time coordinate $(y^0,\bbox{y})$.
It is nothing but the time derivative of the 
Fourier transform of $\Delta_t$ at point
$(t,\bbox{x})$. Obviously a violation of the locality axiom
implies, that this electric field may be measured already for times
$t < \bar{x}$, and therefore it propagates faster than light (FTL).

According to our calculation such a violation may happen when
plasmon propagation and collective effects are treated separately,
i.e., when momenta are restricted to the timelike or spacelike region.
Consequently such a separation should be done very carefully.

However, FTL propagation may also arise with propagators
that are ``patched'' together: Resummed propagators for soft momenta,
and free propagators for hard momenta. Our conclusion is,
that to preserve causality one needs a spectral function which
interpolates in a ``smooth analytical way''
between the high-momentum and the soft-momentum
region, i.e., a naive ``patching'' may lead to unphysical results.

The third conclusion is associated with the separation into transverse
and longitudinal degrees of freedom. From the standard literature one
may get the impression, that they may be treated independently. In
several applications of the propagators (\ref{hqed}), one of the 
two is replaced by a free propagator. As we have argued, this is
an invalid approximation: In order to preserve causality, longitudinal
and transverse propagator must coincide in the limit $|\bbox{p}|=0$.

Let us finally discuss a recipe to obtain local spectral functions.
As noted before, the most general such
function at finite temperature has been given in ref. \cite{BB92,BB94},
where it was also shown that in principle an exponential falloff is
necessary to obey the relativistic KMS-condition.
For any non-local approximation (like e.g. obtained by ``patching''
propagators together) one may proceed as follows.
In coordinate space, the non-local commutator function is multiplied by
$\Theta(t^2-\bar{x}^2)$, then transformed back into momentum space.
Equivalently, one may convolute the old momentum space propagator
with the Fourier transform of such a $\Theta$-function.

We are currently exploring, how such a prescription would affect
the ``patched'' spectral functions of hot gauge theory.
Preliminary calculations show, that indeed this procedure leads
to a spectral function which does not differ too much from
eqn. (\ref{hqed}), but which is nonzero for all values of 
the real energy parameter.
\nopagebreak

\subsection*{Acknowledgements}
We thank D.Buchholz for alerting us to the possible violation
of locality when using approximate spectral functions.
One of us (P.A.H.) wishes to express his gratitude to the members
of the Service Physique Theorique in Saclay for their kind hospitality. 

Gratefully acknowledged are stimulating discussions of this
work with H.A.Weldon, R.Pisarski, L.McLerran and J.P.Blaizot on occasion
of the VIIth Max-Born Symposium in Karpacz/Poland.
Finally, thanks to J.Lindner for his comments on
the manuscript.
\begin{figure}[t]
\vspace*{85mm}
\includegraphics{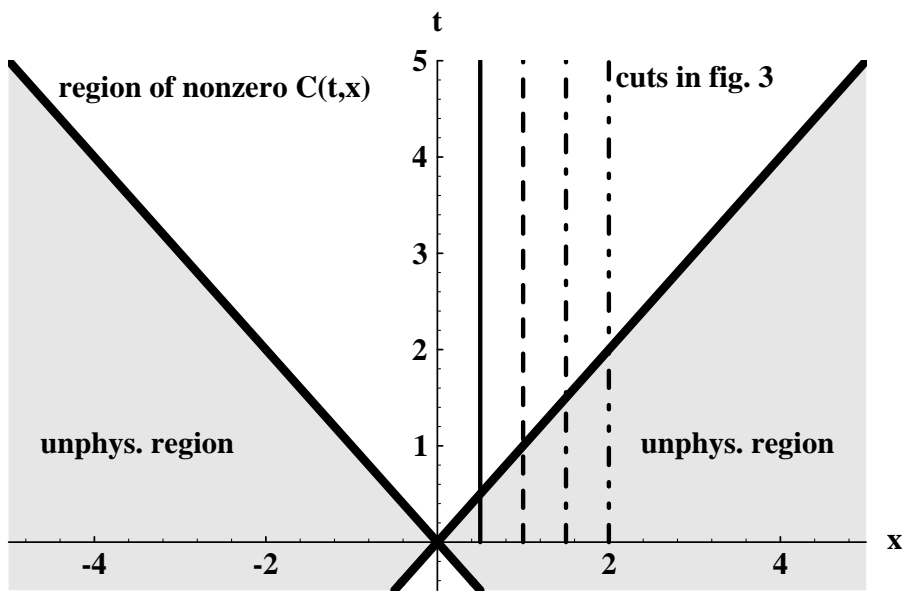}
\caption{Domain of support of a physical (anti-) commutator function
in coordinate space (Unphysical region = shaded area).}
\end{figure}
\begin{figure}[t]
\vspace*{185mm}
\includegraphics{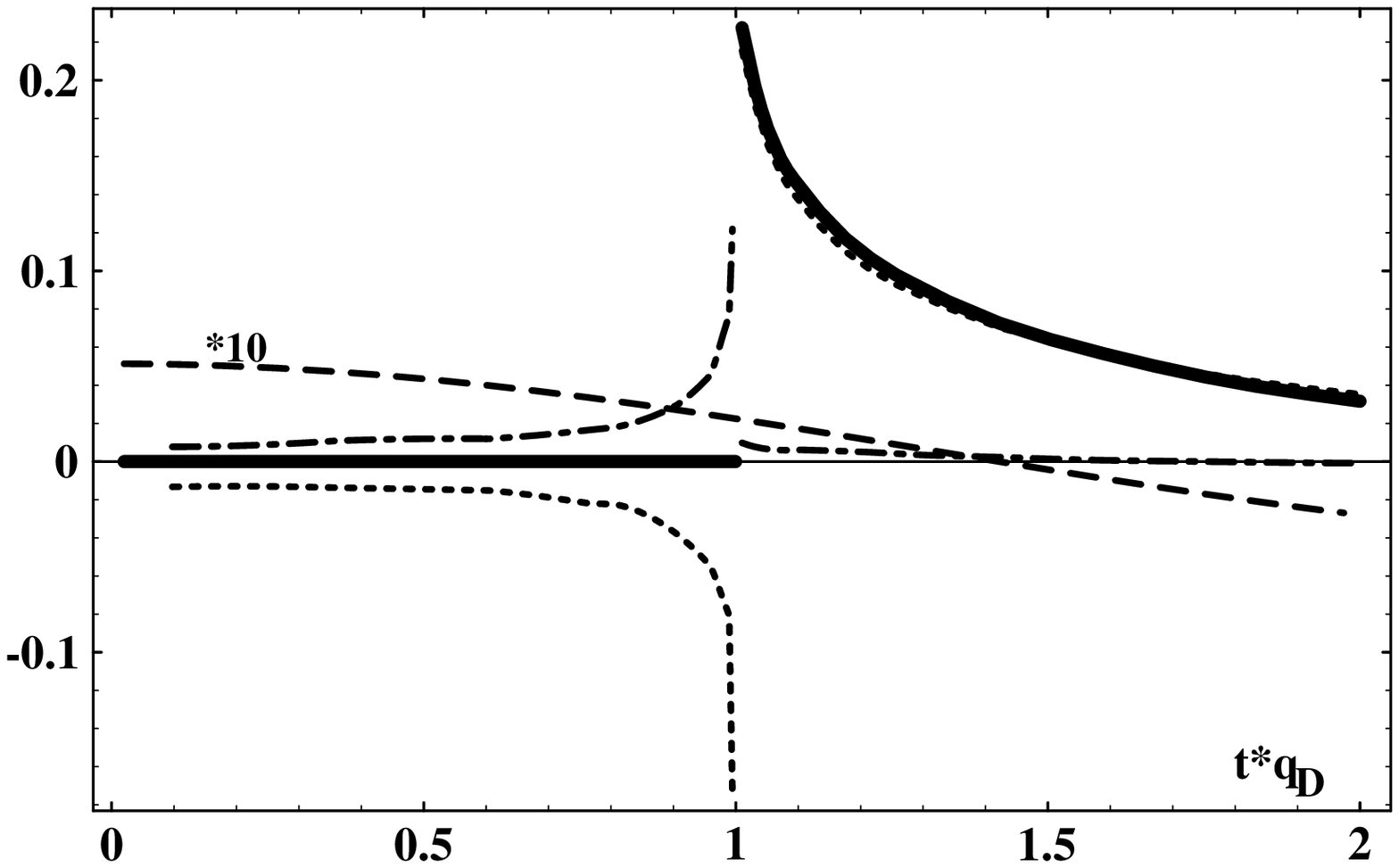}
\includegraphics{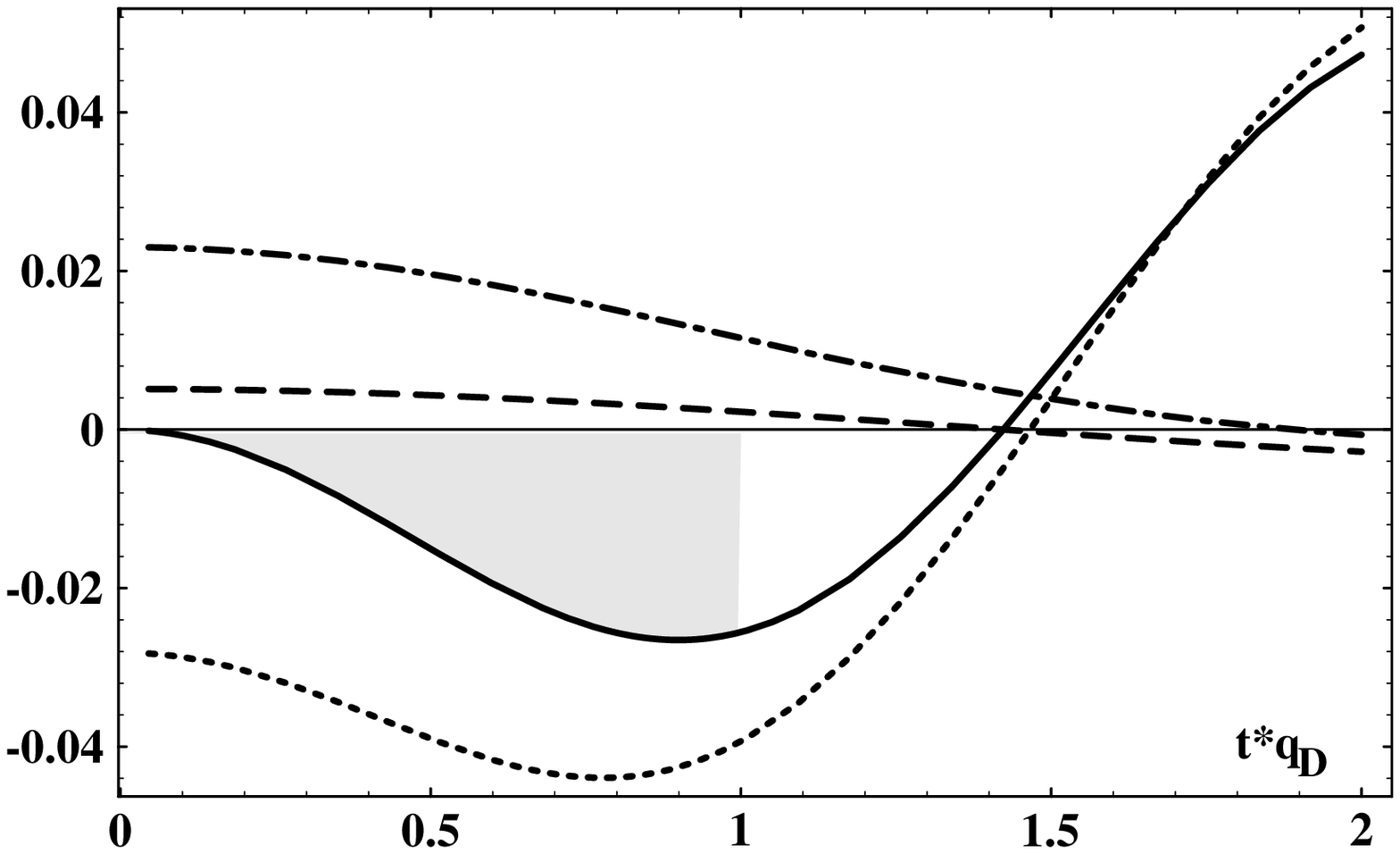}
\caption{Fourier transform of the hot gauge theory fermion spectral 
function at $\bar{x} M_D = 1$.}
{\small
Top panel: resummed propagator (\ref{sfer}) used for {\em all\/} momenta;\\
bottom panel: (\ref{sfer}) for momenta $\bar{p}<\sqrt{8}M_D$, 
 otherwise free propagator.\\
Vertical lines at discontinuity omitted in the top panel.\\
Dotted line: ``particle'' contribution, dashed line ``hole''
contribution, dash-dotted line continuum contribution (see text and ref. 
\cite{W89}).\\
Continuous line: Sum of the three pieces, unphysical contribution shaded.}
\end{figure}
\begin{figure}[t]
\vspace*{185mm}
\includegraphics{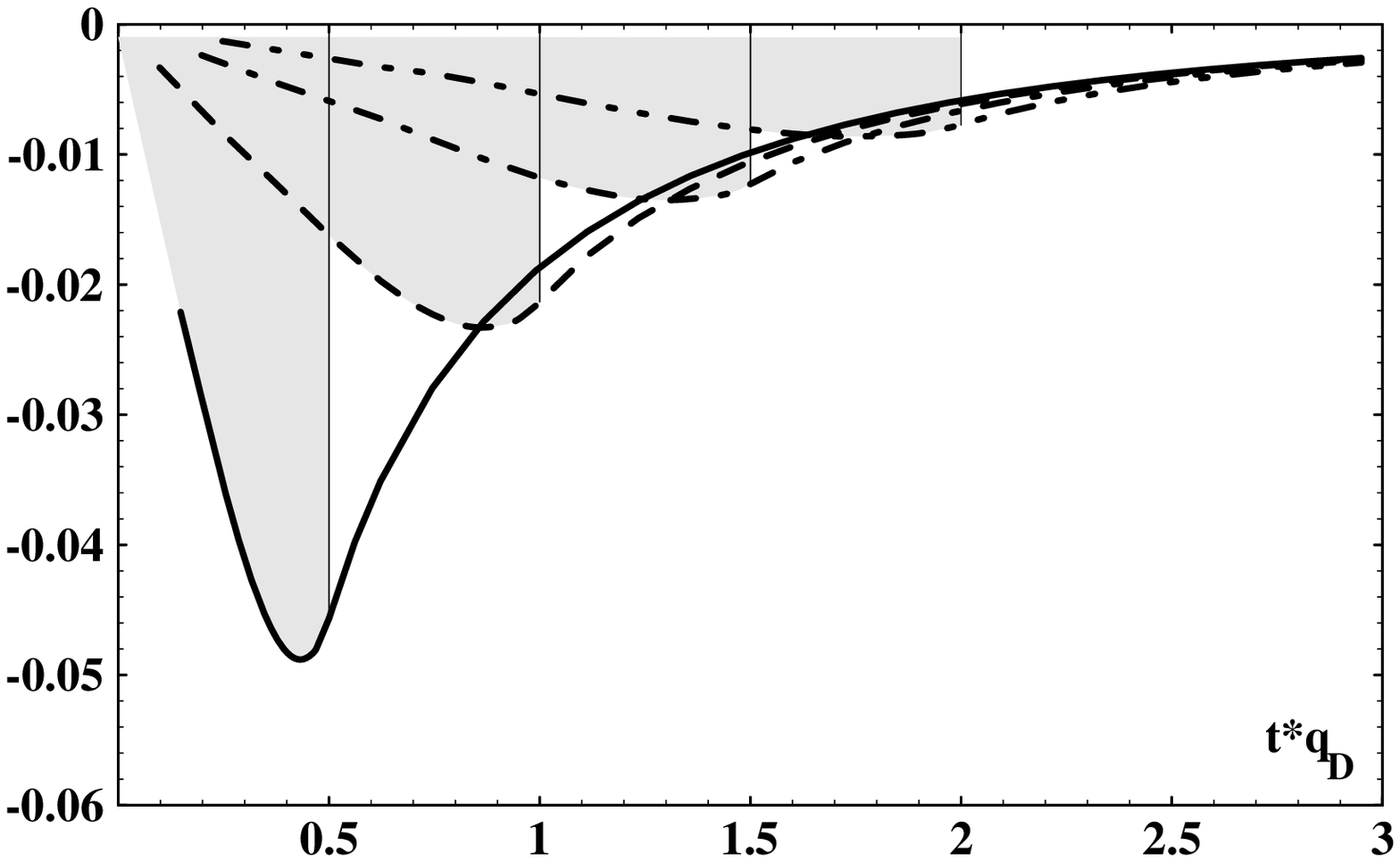}
\includegraphics{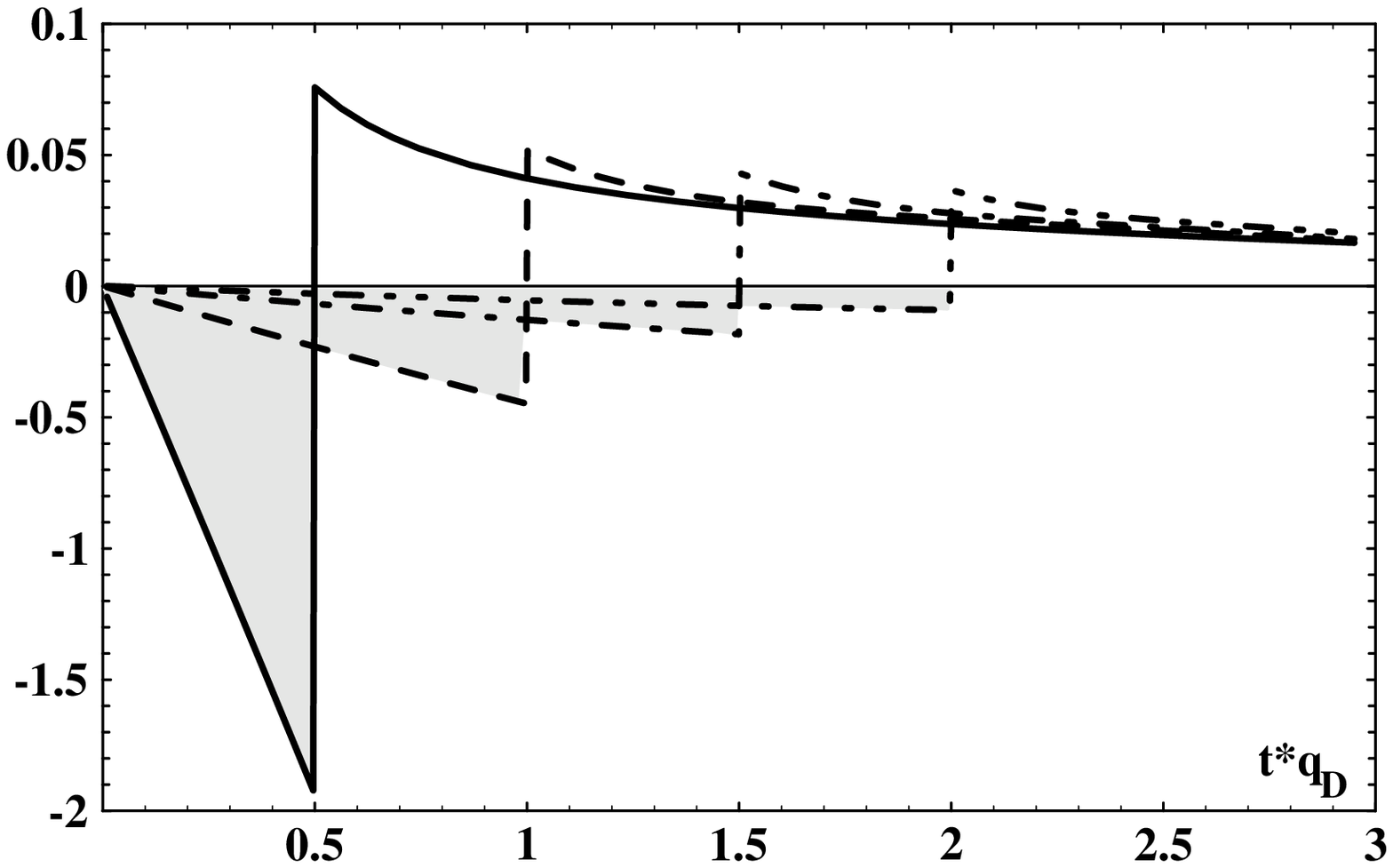}
\caption{Fourier transform of the transverse piece in eq. (\ref{d33}).}
{\small
Top panel: continuous part (spacelike momenta);
bottom panel: plasmon part (timelike momenta),
 $\delta$-functions at $t=\bar{x}$ removed. Note the two
different vertical scales in the bottom panel.\\[1mm]
Plotted for $\bar{x} q_D$ = 0.5 (continuous), 1.0 (dashed), 1.5 (dash-dotted)
and 2.0 (dash-double-dotted).
Contributions outside the forward lightcone are shaded.}
\end{figure}
\end{document}